\documentclass[twocolumn, secnumarabic,amssymb, nobibnotes, aps, prd]{revtex4-2}
\usepackage{hyperref}
\usepackage{amsmath}
\usepackage{graphicx}
\usepackage{float} 
\usepackage{placeins}

\begin{document}
	
	\title{Relic Density Topology as a Discriminatory Tool: A Comparative Analysis of IDM, MSSM, and NMSSM Dark Matter}
	
	\author{Mohid Farhan}
	\email{mohidf35@gmail.com} 
	\affiliation{
		Department of Space Science,
		Institute of Space Technology,
		Islamabad, Pakistan
	}
	
	\date{\today}
	
\begin{abstract}
	This study proposes a diagnostic mechanism based on the \textit{relic density topology} to discriminate between the Inert Doublet Model (IDM), the Minimal Supersymmetric Standard Model (MSSM), and the Next-to-Minimal Supersymmetric Standard Model (NMSSM). Using a unified numerical scan with \texttt{micrOMEGAs} over the heavy-mass regime ($m_{\chi} > 300~\text{GeV}$), we contrast the phenomenological profiles of these frameworks. We demonstrate that the IDM admits a broad, stable ``viability plateau'' driven by efficient gauge couplings, while the MSSM and NMSSM typically overproduce dark matter, reaching the Planck relic density only through narrow, fine-tuned resonance channels. A quantitative fine-tuning measure reveals the IDM's viable parameter space is an order of magnitude more natural than its SUSY counterparts. Furthermore, by examining the thermal decoupling epoch ($z_f$) and the CMB energy-injection parameter ($p_{\rm ann}$), we confirm that all identified viable regions are consistent with cosmological observations, and that the models exhibit different thermal history scenarios for the Universe. Our findings establish a multi-faceted discriminative framework: the IDM is characterized by a robust plateau and low fine-tuning, the MSSM by a sharp slepton-mediated annihilation dip, and the NMSSM by a diluted signature due to singlino admixture. The discovery of a heavy WIMP without sharp resonance features would therefore phenomenologically favor scalar doublet extensions over minimal supersymmetric frameworks.
\end{abstract}
	
	\maketitle

\section{Introduction}

In 1937, Fritz Zwicky observed that objects far from the centers of their galaxies move faster than predicted by Newtonian mechanics \cite{Zwicky1937}. This discrepancy was later confirmed through measurements of galactic rotation curves \cite{galactic_dm, Rubin1980}, providing strong evidence for the existence of non-luminous matter, termed ``dark matter.'' Numerous Beyond the Standard Model (BSM) frameworks have since been proposed to explain this anomaly by introducing new particles, including the Inert Doublet Model (IDM) \cite{idm_gustafsson}, the Minimal Supersymmetric Standard Model (MSSM) \cite{dawson_mssm}, and the Next-to-Minimal Supersymmetric Standard Model (NMSSM) \cite{nmssm_physics}. While many other extensions exist, this study focuses on these three foundational models.

This article proposes and demonstrates a relic-density-based discriminatory mechanism to distinguish between two broad classes of dark matter candidates: those arising from doublet extensions (exemplified by the IDM) and those from supersymmetric frameworks (represented by the MSSM and NMSSM). We show that the topology of the relic density curve, the degree of parametric fine-tuning, and the associated thermal history of the dark sector collectively serve as effective discriminators between these paradigms. The degree of parametric fine-tuning is a long-standing criterion for model naturalness \cite{Barbieri1987} which is measured in this article via a logarithmic sensitivity. Using the observational bounds from Planck data \cite{planck}, we first identify viable dark matter mass regions for each model. We then subject these regions to a suite of consistency tests, including a quantitative fine-tuning and constraints from cosmic microwave background (CMB) energy injection \cite{planck}. These regions are further constrained by indirect detection limits from gamma-ray observations of dwarf spheroidal galaxies \cite{FermiLATdSph}.

Although comparisons between the two SUSY models exist in the literature \cite{porto_mssm_nmssm}, and IDM has been compared to the Inert Triplet Model \cite{IDMvsITM}, a systematic cross-model analysis that compares doublet extensions with IDM is, to our knowledge, absent in literature. By contrasting the relic density behavior across these representative models, this work provides a clearer phenomenological distinction between doublet and SUSY dark matter scenarios. The comprehensive comparison presented here offers a practical reference for model selection in future new-physics searches, especially in light of upcoming cosmological and collider data.
	
	\section{Theoretical Framework}
	\subsection*{Inert Doublet Model (IDM)}
IDM is a doublet extension to the standard model (SM), as it hypothesizes an inert doublet, where the lightest scalar ($H_1$) is the dark matter candidate \cite{Deshpande1978,idm_gustafsson}. "Inert" implies that the doublet does not gain a Vacuum Expectation Value (VEV), and is therefore, unable to participate in Electro-Weak Symmetry Breaking (EWSB) at the tree-level. The doublets are represented by:
	\begin{equation}
	\phi_1 = \begin{pmatrix}
		G^{+} \\
		\frac{v + h + \iota G^0}{\sqrt{2}}
	\end{pmatrix},\phi_2 = \begin{pmatrix}
	H_1^{+} \\
	\frac{H_1^0 + \iota A_1^0}{\sqrt{2}}
	\end{pmatrix}
\end{equation} 
Each doublet contains a charged and neutral component. \(\phi_1\) is the active Higgs doublet in the SM, as indicated by the nonzero vacuum expectation value \(v\), which is responsible for electroweak symmetry breaking via the Higgs mechanism. It is through this mechanism that the Standard Model particles acquire mass. The physical Higgs boson is represented by \(h\), with an experimentally measured mass of about 125 GeV. The fields \(G^+\) and \(G^0\) are the Goldstone bosons. 
On the other hand, a $Z_2$ symmetry is imposed on $\phi_2$ to make the dark matter candidate stable and inert, where the inert doublet is odd under $Z_2$ transformations: 
$$\phi \rightarrow -\phi$$
and the SM particles are even under $Z_2$ transformations:
$$\phi \rightarrow \phi$$
$H_1^0$ is the dark matter candidate (CP-even scalar) and $A_1^0$ is the CP-odd scalar.
The Lagrangian of IDM is the sum of the Standard Model Lagrangian terms and terms involving an additional scalar doublet $\phi_2$ that is odd under a $Z_2$ symmetry:
\begin{align}
	\mathcal{L}_{\text{IDM}} &= \mathcal{L}_{\text{SM}} + (D_\mu \phi_2)^\dagger (D^\mu \phi_2) - V(\phi_1, \phi_2)
\end{align}

The scalar potential is:
\begin{align}
	V(\phi_1, \phi_2) &= \mu_1^2 |\phi_1|^2 + \mu_2^2 |\phi_2|^2 \nonumber \\
	&\quad + \lambda_1 |\phi_1|^4 + \lambda_2 |\phi_2|^4 + \lambda_3 |\phi_1|^2 |\phi_2|^2 \nonumber \\
	&\quad + \lambda_4 |\phi_1^\dagger \phi_2|^2 + \left[ \frac{\lambda_5}{2} (\phi_1^\dagger \phi_2)^2 + \text{h.c.} \right]
\end{align}
The coupling constant relevant to us is $\lambda_L$ which is given by: 
\begin{equation}
	\lambda_L=\frac{1}{2}(\lambda_3+\lambda_4+\lambda_5)
\end{equation}
which depicts the interaction of the Dark Matter (DM) candidate with the Higgs boson. The IDM parameter space is severely constrained by vacuum stability, perturbativity, and most recently, the measured properties of the 125 GeV Higgs boson at the LHC \cite{IDMHiggsLHC}. The mass-deficit between the CP-even and odd candidates within the inert doublet is given by $\delta$. Relic density dependence on $\lambda_L$ with varying DM mass ($m_{DM}$) and $\delta$ is discussed later.
  
\subsection*{Minimal Supersymmetric Standard Model (MSSM)}
The MSSM postulates that every fermionic SM particle has a supersymmetric (SUSY), bosonic counterpart and vice versa. The SUSY particles can be distinguished from SM particles by R-parity operator, which is depicted by: 
\begin{equation}
\hat{R}=(-1)^{3(B-L)+2s}
\end{equation}
where B and L are Baryon and Lepton numbers and s shows the spin state. All SM and SUSY particle return eigenvalues of 1 and -1, respectively under the R-parity operator. This conservation ensures the stability of the Lightest SUSY particle (LSP) or neutralino ($\tilde{\chi}_1^0$), which is the dark matter candidate. Unlike IDM, the mass of the dark matter candidate is a derived quantity that is computed by diagonalizing the mass matrix \cite{dawson_mssm}. The neutralino mass matrix in the MSSM, expressed in the basis 
$(\tilde{B}, \tilde{W}^0, \tilde{H}_d^0, \tilde{H}_u^0)$, determines the composition of the LSP and its interactions. 
The full expression is well-established \cite{dawson_mssm}. The Bino ($\tilde{B}$), Wino ($\tilde{W}^0$) and Higgsinos ($\tilde{H}_d^0, \tilde{H}_u^0$) mixings ultimately determine the mass and properties of dark matter candidate \cite{MSSMCoAnn}. 
\begin{equation}
	\tilde{\chi}_1^0=N_{11}\tilde{B}+N_{12}\tilde{W}^0+N_{13}\tilde{H}_d^0+N_{14}\tilde{H}_u^0
\end{equation}
The crucial parameters are $M_1$ (Bino mass), $M_2$ (Wino mass), $\mu$ (Higgsino mixing) and $\tan\beta$, which is the ratio of VEVs for the up-type to the down-type Higgs doublets. $\mu$ appears in the superpotential:
\begin{equation}
	W_{MSSM}= W_{Yukawa} +\mu H_u H_d
\end{equation}
The effects of $\mu$ and $\tan\beta$ on relic density will be shown in the results and discussion section. $\mu$ comes with a naturalness issue since its mass is not constrained by any symmetry, which means that it can be as large as the Planck Mass and so naturalness becomes a problem.
 	\subsection*{Next-to-Minimal Supersymmetric Standard Model (NMSSM)}
The NMSSM is a singlet extension of the MSSM, naturally solves the '$\mu$ problem'. A gauge singlet superfield $\hat{S}$ is introduced in addition to the two Higgs doublets. This introduces new terms to the superpotential \cite{NMSSM}:
\begin{equation}
	W_{NMSSM}=W_{Yukawa}+\lambda\hat{S}\hat{H_u}\hat{H_d}+\frac{1}{3}\kappa\hat{S}^3
\end{equation}
The mass matrix is now a basis of $(\tilde{B}, \tilde{W}^0, \tilde{H}_d^0, \tilde{H}_u^0,\hat{S})$ \cite{nmssm_physics,Ellwanger2009}
This removes $\mu$  from the superpotential, which naturally solves the $\mu$ problem, while introducing a richer Higgs and neutralino spectrum. It also opens up new annihilation channels and offers more flexibility. In the SUSY models, the lightest supersymmetric particle (LSP) is often considered the dark matter candidate since R-parity forces it to remain inert. While solving the $\mu$-problem, the NMSSM introduces new constraints from searches for light Higgs bosons and exotic decays at the LHC \cite{NMSSMLHC}.
	\section{Methodology}
The relic density $\Omega h^2$ of the dark matter candidate is computed using micrOMEGAs, a numerical package designed to evaluate dark matter observables across BSM frameworks. For each model, the particle content and interactions are specified via model files in CalcHEP format \cite{micromegas,Micromegas5}. The time evolution of the dark matter number density $n_\chi$ is governed by the Boltzmann equation, which micrOMEGAs solves numerically:
\begin{equation}
	\frac{dn_\chi}{dt} + 3Hn_\chi = -\langle \sigma v \rangle \left(n_\chi^2 - n_{\chi}^{\text{eq}\,2} \right),
\end{equation}
where $H$ is the Hubble parameter, $\langle \sigma v \rangle$ is the thermally averaged annihilation cross-section, and $n_\chi^{\text{eq}}$ is the equilibrium number density.
MicrOMEGAs computes the thermally averaged cross-section using:
\begin{equation}
	\langle \sigma v \rangle = \frac{1}{8m_\chi^4 T K_2^2(m_\chi/T)} \int_{4m_\chi^2}^{\infty} ds\, \sigma(s)(s - 4m_\chi^2)\sqrt{s}\,K_1(\sqrt{s}/T),
\end{equation}
where $K_1$ and $K_2$ are modified Bessel functions, and $\sigma(s)$ includes all $2 \to 2$ annihilation and coannihilation channels.
The relic abundance is derived via numerical integration over temperature:
\begin{equation}
	\Omega_\chi h^2 = \frac{1.07 \times 10^9\,\text{GeV}^{-1}}{g_*^{1/2} M_{\text{Pl}} \int_{x_f}^{\infty} \frac{\langle \sigma v \rangle}{x^2} dx},
\end{equation}

where $x = m_\chi/T$, $x_f$ is the freeze-out point, $g_*$ denotes effective relativistic degrees of freedom, and $M_{\text{Pl}}$ is the Planck mass. The relic density of dark matter is constrained to be in the region: 
\begin{equation}
	\Omega h^2 =0.1199 \pm 0.0027
\end{equation} 
by Planck data \cite{planck}. To maintain consistency with observation, we will include these bounds in our computational analysis. 

All relevant parameters are given in TABLE I. For each model, the dark matter mass parameter ($m_{H_1}$ in the IDM and $M_1$ in the MSSM/NMSSM) was scanned over the ranges shown in Table I, with a stepsize of 1 GeV, while all other parameters were fixed to the values listed. The parameter files provided to micrOMEGAs serve as benchmark inputs and are programmatically updated during the scan.

The direct comparison of dark matter relic density also allows us to compare fine-tuning in viability regions as dictated by Eq. 12. To quantify the parametric sensitivity of the relic density to variations in the dark matter mass, we define a logarithmic fine-tuning measure similar in spirit to \cite{Anderson1995},
\begin{equation}
	\Delta_{M_{\rm DM}} \equiv 
	\left|
	\frac{\partial \ln \Omega h^2}{\partial \ln M_{\rm DM}}
	\right|,
\end{equation}
which characterizes the relative change in the relic abundance under infinitesimal variations of the dark matter mass. Our definition of $\Delta_{M_{\text{DM}}}$ quantifies the pointwise logarithmic sensitivity of the relic density to variations in the dark matter mass. While not equivalent to the multi-parameter Barbieri--Giudice measure~\cite{Barbieri1987}, which assesses naturalness across the entire parameter space, $\Delta_{M_{\text{DM}}}$ serves as a model-independent comparative tool for the specific sensitivity relevant to relic density predictions since $m_{DM}$ is the only truly comparable parameter in all models. It directly captures how precisely the dark matter mass must be tuned to achieve the observed $\Omega h^2$, which is particularly relevant for distinguishing broad plateaus (low $\Delta$) from narrow resonances (high $\Delta$).

The derivative is evaluated via central finite differences with a fixed mass step $\Delta M = 1$ GeV. This step is small compared to the typical mass scales in our scan (300--1000~GeV) and is, at worst, comparable to SUSY resonance peaks. We have verified the numerical stability of this approximation by comparing with smaller step sizes (0.5~GeV) at selected points near resonances, finding variations in $\Delta$ of less than $10\%$. For the extremely steep MSSM resonances, the computed $\Delta$ values are large (up to $\sim 90$), reflecting genuine physical sensitivity rather than numerical artifacts. We note that these resonances are resolved by multiple scan points (Table II), confirming that our step size is adequate to capture their characteristic fine-tuning. 

We employ a central difference approximation,
\begin{equation}
	\left.
	\frac{\partial \ln \Omega h^2}{\partial \ln M_{\rm DM}}
	\right|_{i}
	\;\approx\;
	\frac{
		\ln\!\left(\Omega_{i+1} h^2\right) - \ln\!\left(\Omega_{i-1} h^2\right)
	}{
		\ln M_{i+1} - \ln M_{i-1}
	},
\end{equation}
which provides a stable numerical estimate of the logarithmic sensitivity and is insensitive to the absolute scale of the mass parameter.

The fine-tuning measure is evaluated exclusively for parameter points satisfying the Planck 2018 constraint on the dark matter relic density (Eq. 12), thereby ensuring that the sensitivity is assessed only within phenomenologically viable regions of parameter space.

In cases where the Planck-allowed parameter space is fragmented into multiple disconnected regions, the fine-tuning measure is computed separately within each region. This procedure avoids numerical artifacts that would otherwise arise from finite differencing across gaps in the parameter scan. The same fine-tuning prescription is applied consistently to the Inert Doublet Model (IDM), the MSSM, and the NMSSM, enabling a meaningful comparison of the degree of mass sensitivity associated with each framework by using the mean value for all viable points defined by:
$$\langle\Delta\rangle = \frac{1}{N}\sum_{i=1}^{N} |\Delta_i|$$
which automatically rewards models that have a wide viability region (N), with a lower $\langle\Delta\rangle$ corresponding to less fine-tuning. The finite difference method was applied using \textbf{python}.

For each viable parameter point satisfying the relic density constraint, we extract additional freeze-out–related quantities characterizing the thermal decoupling of dark matter.

The freeze-out parameter is defined as
\begin{equation}
	X_f \equiv \frac{m_{\rm DM}}{T_f},
\end{equation}
where $m_{\rm DM}$ is the dark matter mass and $T_f$ is the freeze-out temperature.  
The value of $X_f$ is obtained directly from \texttt{micrOMEGAs} during the relic density computation.

The freeze-out temperature is then computed as
\begin{equation}
	T_f = \frac{m_{\rm DM}}{X_f}.
\end{equation}

To associate freeze-out with a cosmological epoch, we compute the corresponding redshift using entropy conservation \cite{Kolb1990},
\begin{equation}
	1 + z_f
	= \frac{T_f}{T_0}
	\left( \frac{g_{*S}(T_f)}{g_{*S}(T_0)} \right)^{1/3},
\end{equation}
where $T_0 = 2.35 \times 10^{-13}\,\mathrm{GeV}$ is the present CMB temperature,  
$g_{*S}(T_0)=3.91$ is the present entropy degrees of freedom,  
and $g_{*S}(T_f)=106.75$ is assumed for freeze-out occurring in the relativistic Standard Model plasma. These quantities turn out to be different despite yielding the same relic density, and therefore a comparison can be drawn.

Furthermore, we check the compatibility of all viable regions with the Cosmic Microwave Background (CMB) data \cite{Planck2015}. Energy injection from dark matter annihilation during the epoch of recombination alters the ionization history of the Universe and leaves imprints on the CMB anisotropies. These effects are conveniently captured by the effective annihilation parameter $p_{\mathrm{ann}}$ \cite{Slatyer2015, Madhavacheril2014}, which is constrained by Planck observations.

In this work, $p_{\mathrm{ann}}$ is defined as
\begin{equation}
	p_{\mathrm{ann}} \equiv f_{\mathrm{eff}} \, \frac{\langle \sigma v \rangle_{v \to 0}}{m_{\mathrm{DM}}},
\end{equation}
where $\langle \sigma v \rangle_{v \to 0}$ is the thermally averaged dark matter annihilation cross section in the zero-velocity limit, $m_{\mathrm{DM}}$ is the dark matter mass, and $f_{\mathrm{eff}}$ is the efficiency factor encoding the fraction of injected energy that contributes to ionization and heating of the primordial plasma.

The annihilation cross section $\langle \sigma v \rangle_{v \to 0}$ is computed using \texttt{micrOMEGAs} via the \texttt{calcSpectrum} routine, which evaluates the annihilation rate at rest and includes all kinematically accessible final states. This quantity is appropriate for CMB constraints, as dark matter particles are non-relativistic at recombination.

Following standard practice in the literature \cite{Slatyer2015}, we adopt a representative constant value
\begin{equation}
	f_{\mathrm{eff}} = 0.3,
\end{equation}
which lies within the range inferred for weak-scale dark matter annihilating dominantly into electroweak final states.

For each parameter point satisfying the Planck relic density constraint, $p_{\mathrm{ann}}$ is evaluated and compared against the Planck CMB bound \cite{Planck2015},
\begin{equation}
	p_{\mathrm{ann}} \lesssim 3.2 \times 10^{-28} \; \mathrm{cm^3\,s^{-1}\,GeV^{-1}},
\end{equation}
thereby assessing the compatibility of the viable dark matter parameter space with CMB observations.

\begin{table*}[t]
	\centering
	\caption{Parameter choices for the heavy mass regime scans ($300 < m_{\chi} < 1000$ GeV). Parameters without explicit units are dimensionless. Fixed values are shown; the scanned parameter defining the dark matter mass scale is indicated for each model.}
	\label{tab:parameters_heavy}
	\renewcommand{\arraystretch}{1.2}
	\begin{tabular}{|l|c|c|c|}
		\hline
		\textbf{Parameter} & \textbf{IDM} & \textbf{MSSM} & \textbf{NMSSM} \\
		\hline\hline
		\textbf{Scanned mass parameter} 
		& $m_{H_1}$ (300--1000 GeV) 
		& $M_1$ (290--1130 GeV) 
		& $M_1$ (290--1110 GeV) \\
		\hline
		$\mu$ or $\mu_{\mathrm{eff}}$ (GeV) & -- & 1000 & 1000 \\
		\hline
		$M_2$ (GeV) & -- & 2000 & 2000 \\
		\hline
		$M_3$ (GeV) & -- & 2000 & 2000 \\
		\hline
		\textbf{Slepton masses} (GeV) & -- & 1200 & 1200$^{\dagger}$ \\
		\hline
		\textbf{Squark masses} (GeV) & -- & 1000 & 1000$^{\dagger}$ \\
		\hline
		$\tan\beta$ & -- & 30 & 30 \\
		\hline
		$m_{h_{\mathrm{SM}}}$ (GeV) & 125 & -- & -- \\
		\hline
		$m_{H^{\pm}}, m_A$ (GeV) & -- & 1000 & -- \\
		\hline
		$\lambda_L$ & 0.001 & -- & -- \\
		\hline
		$\lambda$ & -- & -- & 0.055 \\
		\hline
		$\kappa$ & -- & -- & 0.03 \\
		\hline
		$A_t$ (GeV) & -- & 1000 & 1000 \\
		\hline
		$A_b$ (GeV) & -- & 1000 & 1000 \\
		\hline
		$A_\tau$ (GeV) & -- & 500 & 1000 \\
		\hline
		$A_\lambda$ (GeV) & -- & -- & 100 \\
		\hline
		$A_\kappa$ (GeV) & -- & -- & $-50$ \\
		\hline
		\textbf{Inert scalar splitting} $\delta$ (GeV) & 10$^{*}$ & -- & -- \\
		\hline
	\end{tabular}
	
	\vspace{0.6em}
	{\footnotesize
		$^{*}$ In the IDM, $m_{H_3} = m_{H_C} = m_{H_1} + 10$ GeV to enable co-annihilation.\\
		$^{\dagger}$ For the NMSSM, slepton and squark masses are specified for the second and third generations; first generation masses are set equal to the second by the spectrum generator.
	}
\end{table*}

\section{Results and Discussion}

In this section, we present the dark matter relic density as a function of key model parameters, manifesting in a cross-model comparison using the dark matter candidate mass as the unifying variable. This comparison forms the basis of the proposed discriminatory mechanism between doublet and supersymmetric extensions to the Standard Model. We also compute the fine-tuning metric for each model and provide a comparison of sensitivity of relic density to DM mass. 

\subsection*{Inert Doublet Model (IDM) Behaviour}

The IDM dark matter relic density depends on parameters like $\lambda_L$, $\delta$ (mass deficit between CP-even and CP-odd candidate) and $m_{DM}$, with its topological dependence illustrated in FIG. 1 and FIG. 2.

FIG. 1 shows the relic density for a fixed DM mass of $m_{\text{DM}} \approx 550$ GeV in the IDM as a function of the DM-Higgs coupling $\lambda_L$, for different values of the mass-splitting parameter $\delta$. As $\delta$ approaches zero, inverting the sign of $\lambda_L$ yields identical relic density values while swapping the charge-parity behaviour of the DM candidate. This symmetry arises from the dominance of the $H_1 A_1$ annihilation channel, whose cross-section scales as $\lambda_L^2$. The symmetry is broken for larger $\delta$, as the abundance of $A_1$ is Boltzmann-suppressed \cite{idm_gustafsson} and other channels with linear dependence on $\lambda_L$ become relevant. FIG. 1 also demonstrates that larger $\delta$ values enhance annihilation efficiency, primarily through the increasingly effective $H_1 H_1 \to f\bar{f}$ channel.

\begin{figure}[H]
	\centering
	\includegraphics[width=0.7\linewidth]{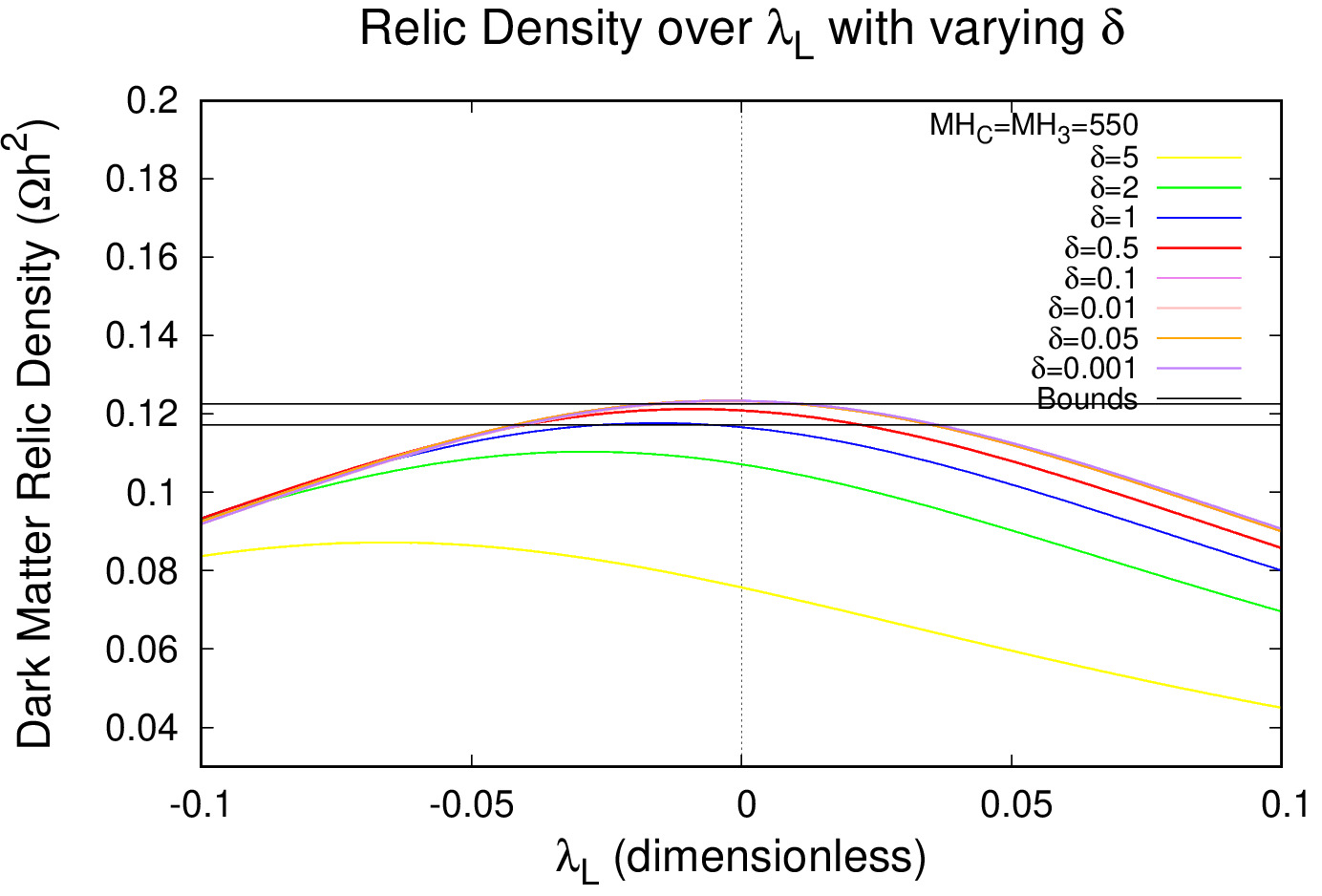}
	\caption{Dark matter relic density in the IDM as a function of the DM-Higgs coupling $\lambda_L$ for varying mass-splitting parameter $\delta$.}
	\label{fig:idm_lambda_delta}
\end{figure}

FIG. 2 confirms that the relic density is highest at low $\lambda_L$, as expected from suppressed annihilation. For the mass range $350 < m_{\text{DM}} < 650$ GeV and with $\delta=0.1$, an increase in DM mass correlates with an increase in relic density. Applying the Planck bounds \cite{planck}, we find a narrow viable mass window of 536 GeV $\leq m_{\text{DM}} \leq$ 548 GeV within this parameter space.

\begin{figure}[H]
	\centering
	\includegraphics[width=0.7\linewidth]{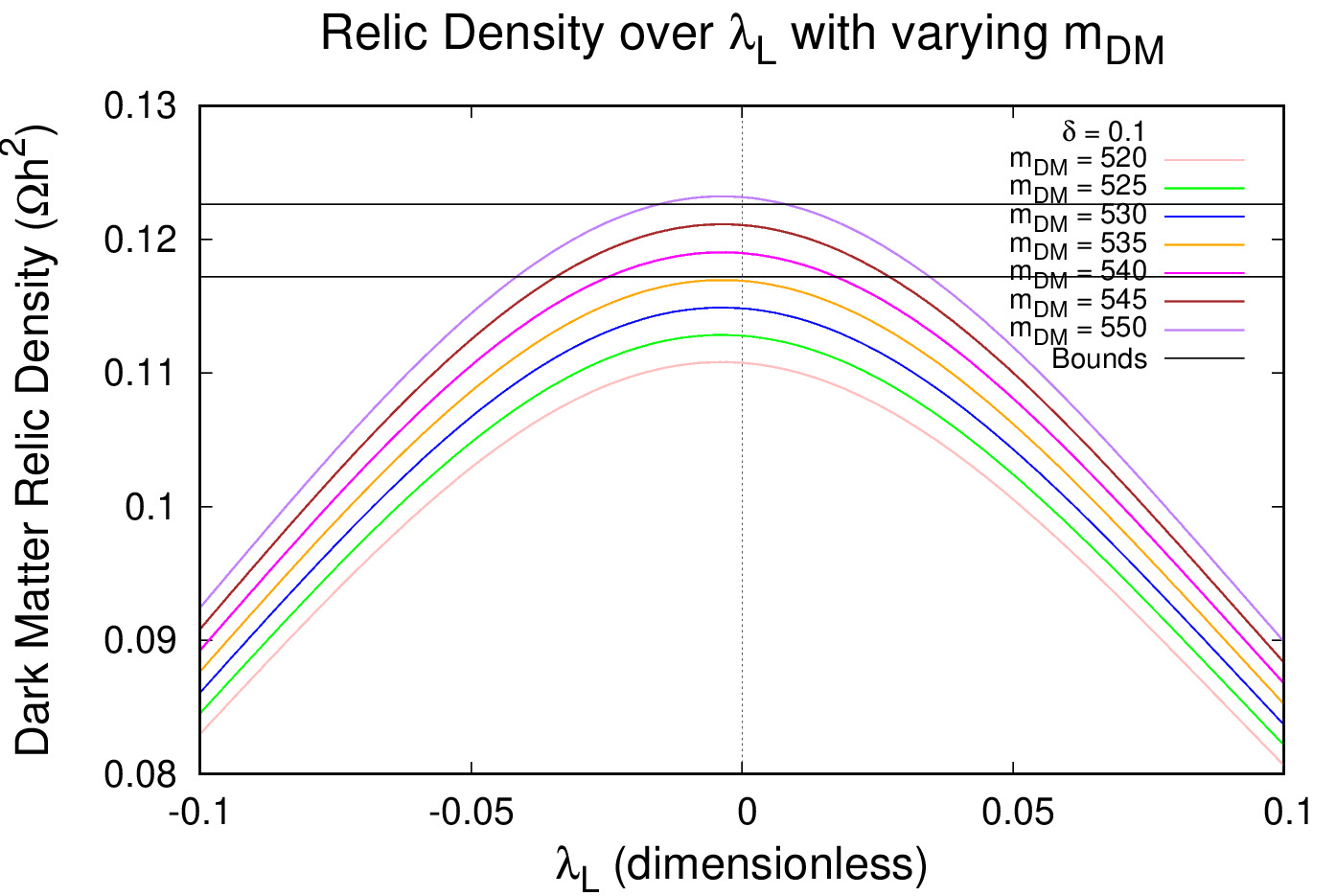}
	\caption{Relic density in the IDM as a function of DM mass for different $\lambda_L$, with $\delta=0.1$. The horizontal band indicates the Planck observational bounds.}
	\label{fig:idm_mass_scan}
\end{figure}

In the heavy mass regime analysis, all parameters will be kept constant with values shown in Table I, with an incremental increase in $m_{DM}$, setting up the topological discrimination of its profile with SUSY models.

\subsection*{Minimal Supersymmetric Standard Model (MSSM) Behavior}

This section demonstrates the topological dependence of the MSSM relic density on key parameters governing neutralino composition: the Higgsino mass parameter $\mu$, the Wino mass parameter $M_2$, and $\tan\beta$. The goal is to illustrate how relic density curves vary with changes in the dominant mass component of the Lightest Supersymmetric Particle (LSP).

\begin{figure}[H]
	\centering
	\includegraphics[width=0.7\linewidth]{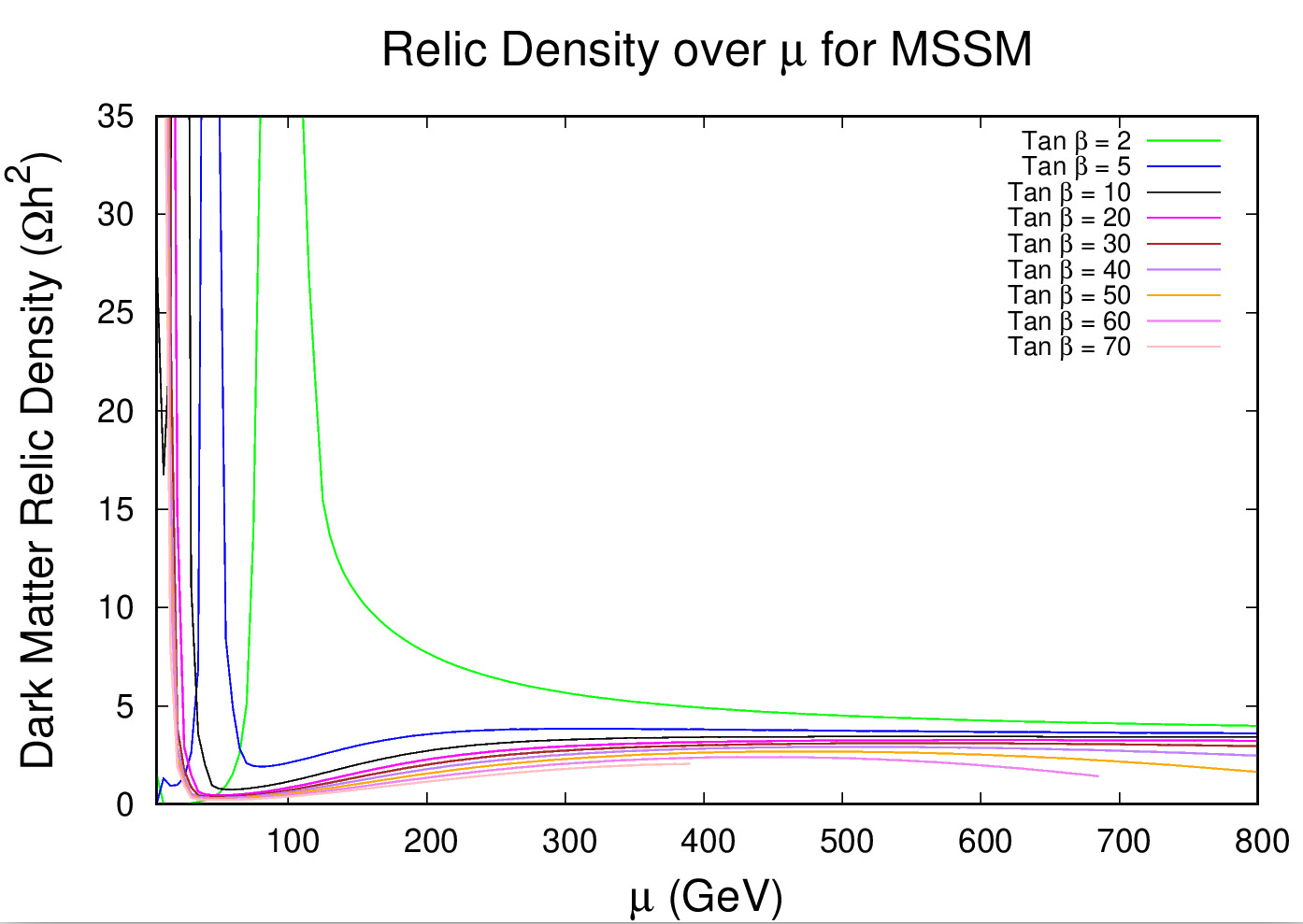}
	\caption{Relic density in the MSSM as a function of $\mu$ and $\tan\beta$, illustrating resonance peaks and the dependence on LSP composition.}
	\label{fig:mssm_mu_tanb}

\end{figure}

FIG. 3 shows $\Omega h^2$ as a function of $\mu$ and $\tan\beta$ while keeping $M_1$ and $M_2$ fixed at large values ($2$ TeV) to isolate a Bino-like LSP. Resonance peaks appear for $\mu \lesssim 100$ GeV, where the LSP becomes Higgsino-like and annihilates efficiently through $s$-channel Higgs or $Z$ exchange. Beyond this region, the LSP transitions to a Bino-like state, and the relic density plateaus. At low $\tan\beta$, suppressed Higgs couplings to down-type quarks result in broader peaks, while higher $\tan\beta$ enhances Yukawa couplings, sharpening the resonances and reducing the overall relic density. Discontinuities in the curves at high $\tan\beta$ and $\mu$ correspond to regions where the SUSY spectrum generator encounters tachyonic instabilities; these unphysical points are excluded from subsequent analysis.

\begin{figure}[H]
	\centering
	\includegraphics[width=0.7\linewidth]{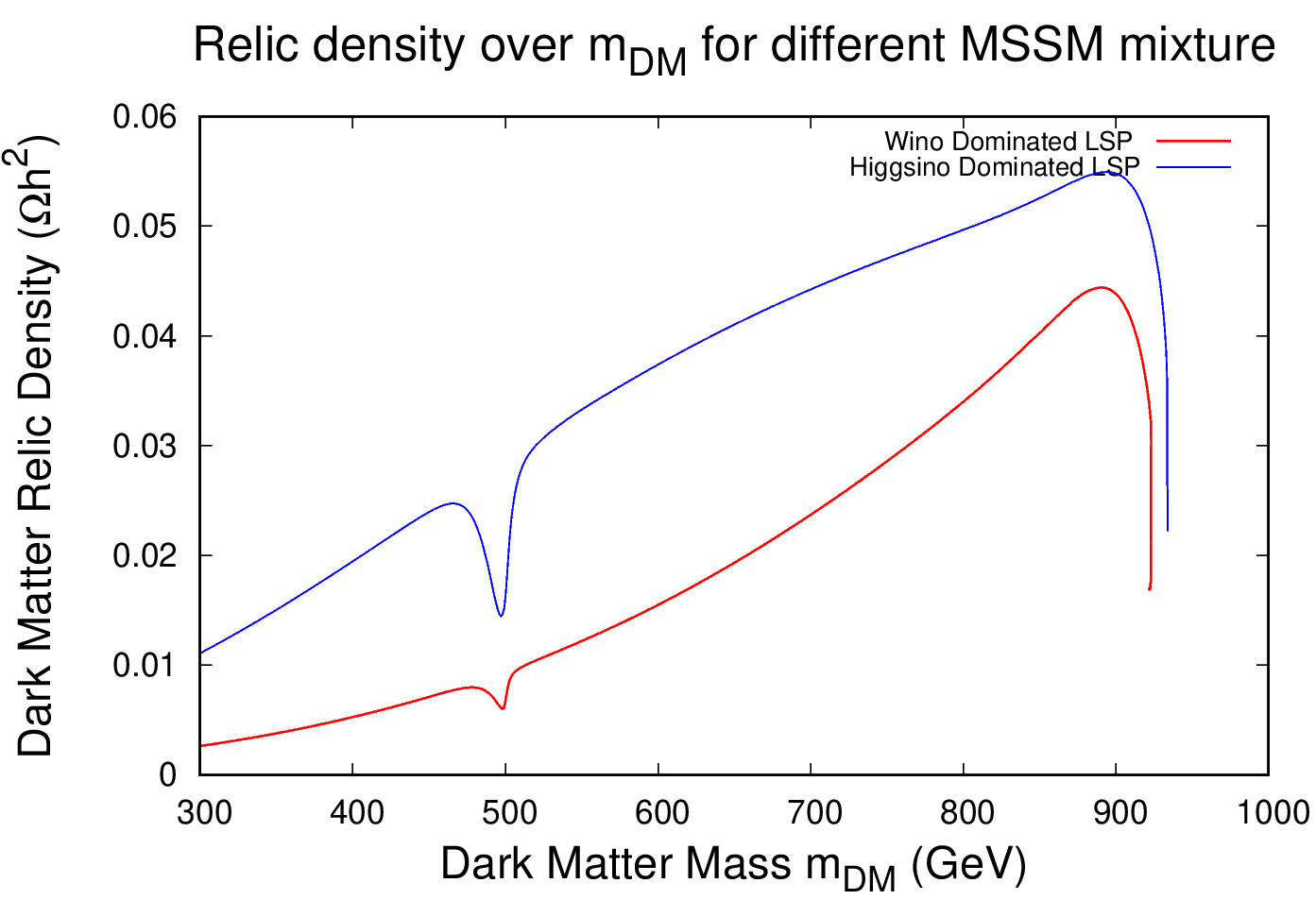}
	\caption{The topological profiles of Wino and Higgsino-like DM in the MSSM.}
	\label{fig:fig4}
\end{figure}

To systematically explore the relic density topology as a function of LSP mass, we vary one neutralino mass parameter at a time while decoupling the others. This yields three characteristic composition regimes: Bino-like ($M_1$ varied, $\mu$, $M_2$ large), Wino-like ($M_2$ varied, $M_1$, $\mu$ large), and Higgsino-like ($\mu$ varied, $M_1$, $M_2$ large). FIG. 4 shows the resulting relic density curves for Wino-like and Higgsino-like LSPs. Both exhibit similar topological profiles—including a characteristic slepton-mediated dip near $500$ GeV—but are shifted downward in normalization due to their more efficient gauge-mediated annihilation.

In the heavy-mass regime ($m_\chi > 300$ GeV) explored in the main comparison (Table I), we fix $\mu = 1$ TeV and $M_2 = 2$ TeV, then vary $M_1$ to produce a Bino-dominated LSP with a mass $m_{\text{LSP}} \approx M_1$. This choice is motivated by two factors:

\textbf{Comparability}: A Bino-like neutralino is a gauge singlet, analogous to the scalar singlet in the IDM, allowing a cleaner comparison of annihilation mechanisms.

\textbf{Phenomenological relevance}: Under our parameter assumptions (Table I), Bino-like LSPs can produce relic densities within the Planck band through specific channels (e.g., slepton-mediated annihilation), whereas pure Wino or Higgsino LSPs typically underproduce dark matter.

Squark and slepton masses are set sufficiently high to ensure that the lightest supersymmetric particle is electrically neutral and colorless, thus avoiding cosmologically disfavored charged or colored dark matter. For the NMSSM, the soft-breaking parameters $A_\kappa$ and $A_\lambda$ are chosen to guarantee successful electroweak symmetry breaking and to avoid tachyonic states in the Higgs sector. The superpotential couplings $\lambda$ and $\kappa$ are taken to be small enough that the singlino component remains subdominant in the lightest neutralino, yet large enough to yield phenomenology distinct from the MSSM. This parameter choice ensures that the dark matter candidate in both SUSY models is predominantly a gauge singlet (Bino-like in the MSSM, Bino-singlino mixed in the NMSSM), facilitating a direct comparison with the scalar singlet of the IDM. All parameter values are summarized in Table I. 

\subsection*{Cross-Model Comparison: Light Mass Regime}
The dark matter candidate mass and its relic density are the primary comparable quantities across all three models. To enable a systematic comparison in the light mass regime (50 GeV $< m_{\text{DM}} <$ 200 GeV), we decouple the Higgsino and Wino sectors by setting $M_2 = 2000$ GeV and $\mu = 1500$ GeV, ensuring $m_{\text{DM}} \approx M_1$. Slepton masses were capped at 400 GeV. The results are shown in FIG. 5.

\begin{figure}[H]
	\centering
	\includegraphics[width=0.7\linewidth]{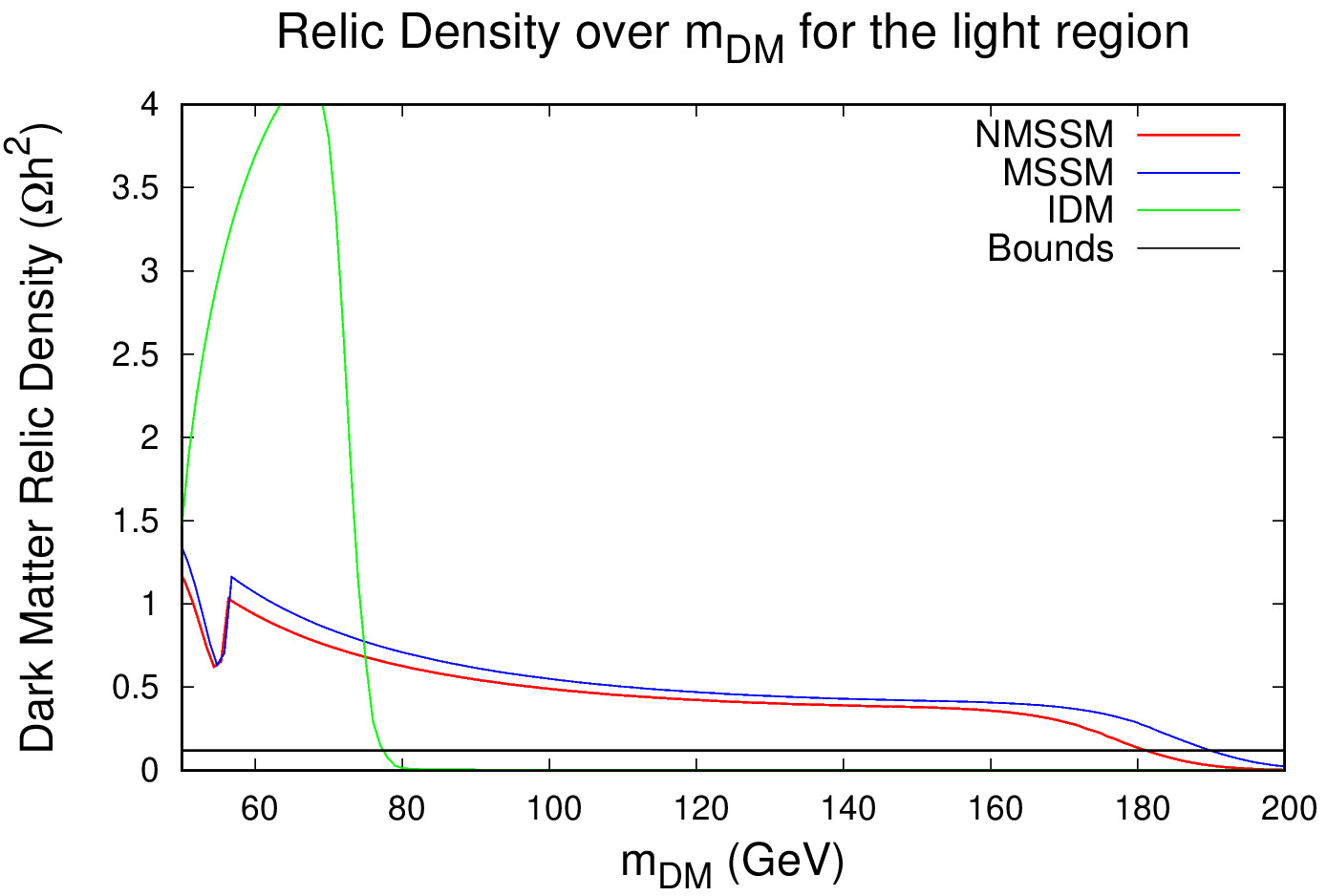}
	\caption{Cross-model comparison of relic densities in the light dark matter mass regime (50–200 GeV) for the IDM, MSSM, and NMSSM. The horizontal band indicates the Planck bounds.}
	\label{fig:comparison_light}
\end{figure}

A prominent Higgs-resonance drop is observed in the IDM at $m_{\text{DM}} \approx M_h/2 = 62.5$ GeV. With large mass splittings to mimic a natural, non-fine-tuned scenario, a viable point is found at $m_{\text{IDM}} \approx 78$ GeV. In contrast, both SUSY models exhibit a gradual decrease in relic density with increasing mass. This trend occurs because the neutralino becomes less Bino-like as $M_1$ increases; even a small Higgsino admixture significantly enhances annihilation, reducing the relic density. The sleptons are kept at least 100–120 GeV heavier than the Bino mass to prevent a charged LSP. With a small coupling $\lambda \approx 0.05$ in the NMSSM to avoid unphysical masses, the SUSY models yield similar densities, with viable points at $m_{\text{MSSM}} \approx 190$ GeV and $m_{\text{NMSSM}} \approx 181$ GeV. While recent constraints disfavor much of this light mass window \cite{LZ2023, XENONnT2023}, narrow viable regions persist near resonances, remaining phenomenologically interesting provided they evade direct detection limits.

\subsection*{Cross-Model Comparison: Heavy Mass Regime}

For the heavy mass regime favored by the IDM ($m_{\rm DM} \gtrsim 300$ GeV), current direct detection bounds are less stringent. The spin-independent cross section in the IDM scales as $\sigma_{\rm SI} \propto \lambda_L^2$, and with $\lambda_L \sim 0.001$ (Table I), we estimate $\sigma_{\rm SI} \sim 10^{-47} - 10^{-46}$ cm$^2$ for the viable mass ranges identified in following section. These values lie within the projected sensitivity of next-generation experiments like DARWIN~\cite{DARWIN}. The broad ``viability plateau'' of the IDM thus presents a well-defined, natural target for future ton-scale detectors, unlike the finely-tuned SUSY resonances which would require precise mass reconstruction.

We next explore the heavy mass regime (300 GeV $< m_{\text{DM}} <$ 1000 GeV), as shown in FIG. 6. Parameters were adjusted to maintain a Bino-dominant LSP ($\mu = 1$ TeV, $M_2 = 2$ TeV), with slepton masses raised to 1.2 TeV. The SUSY models continue their decreasing trend, while the IDM curve remains closer to the experimental bounds.

A key discriminatory feature appears in the MSSM: a distinct resonance dip near $m_{\text{DM}} \approx 500$ GeV. This dip corresponds to the opening of the $t$-channel slepton-mediated annihilation process:
$$
\tilde{\chi}_1^0\, \tilde{\chi}_1^0 \rightarrow l^+\, l^- \quad \text{via } \tilde{l} \text{ (t-channel)}
$$

This channel is highly effective in the MSSM. Its effect is significantly dampened in the NMSSM due to the singlino component of the LSP and the introduction of additional $s$-channel interactions from new CP-even and CP-odd scalars. These competing, less-efficient $s$-channel processes dilute the $t$-channel efficacy, resulting in a higher relic density for the NMSSM in this region. This serves as a clear signature for discriminating between the two SUSY frameworks. 

\begin{figure}[H]
	\centering
	\includegraphics[width=0.65\linewidth]{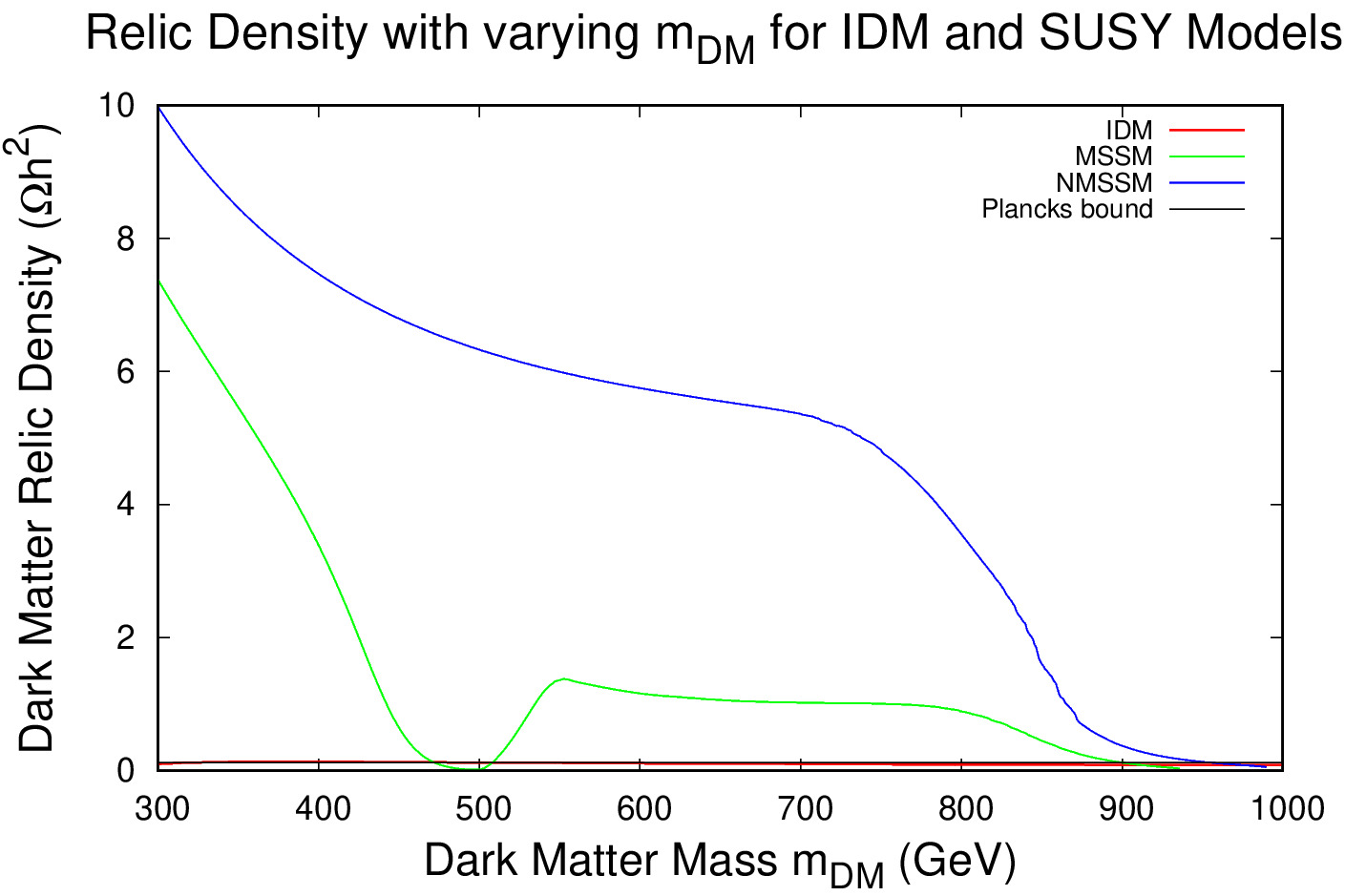}
	\caption{Cross-model comparison of relic densities in the heavy dark matter mass regime (300–1000 GeV). The slepton-mediated dip near 500 GeV in the MSSM is a key discriminating feature.}
	\label{fig:comparison_heavy}
\end{figure}

The pronounced dip at 500 GeV (in this case) is unique to the MSSM in this regime, and its subsequent damping in the NMSSM provide a clear particle-physics fingerprint for distinguishing between the two SUSY frameworks.

\subsection*{Viable Mass Windows and Discriminatory Power}

A zoomed-in view (FIG.7) highlights the viable mass regions consistent with Planck bounds. The IDM accommodates the widest viable regions, 320 GeV $< m_{\text{IDM}} <$ 325 GeV and 382 GeV $< m_{\text{IDM}} <$ 526 GeV. The MSSM yields two narrow viable windows at $m_{\text{MSSM}} \approx 471$ GeV and 508 GeV (driven by the slepton-channel dip), eventually meeting the observational bound near 900 GeV. The NMSSM, however, provides only a single viable point at $m_{\text{NMSSM}} \approx 955$ GeV under these scan conditions. The sharp cut-off in the SUSY curves near 1 TeV occurs as the LSP mass relation $m_{\text{LSP}} \approx M_1$ breaks down due to the growing comparability of slepton and Bino masses.

\begin{figure}[H]
	\centering
	\includegraphics[width=0.65\linewidth]{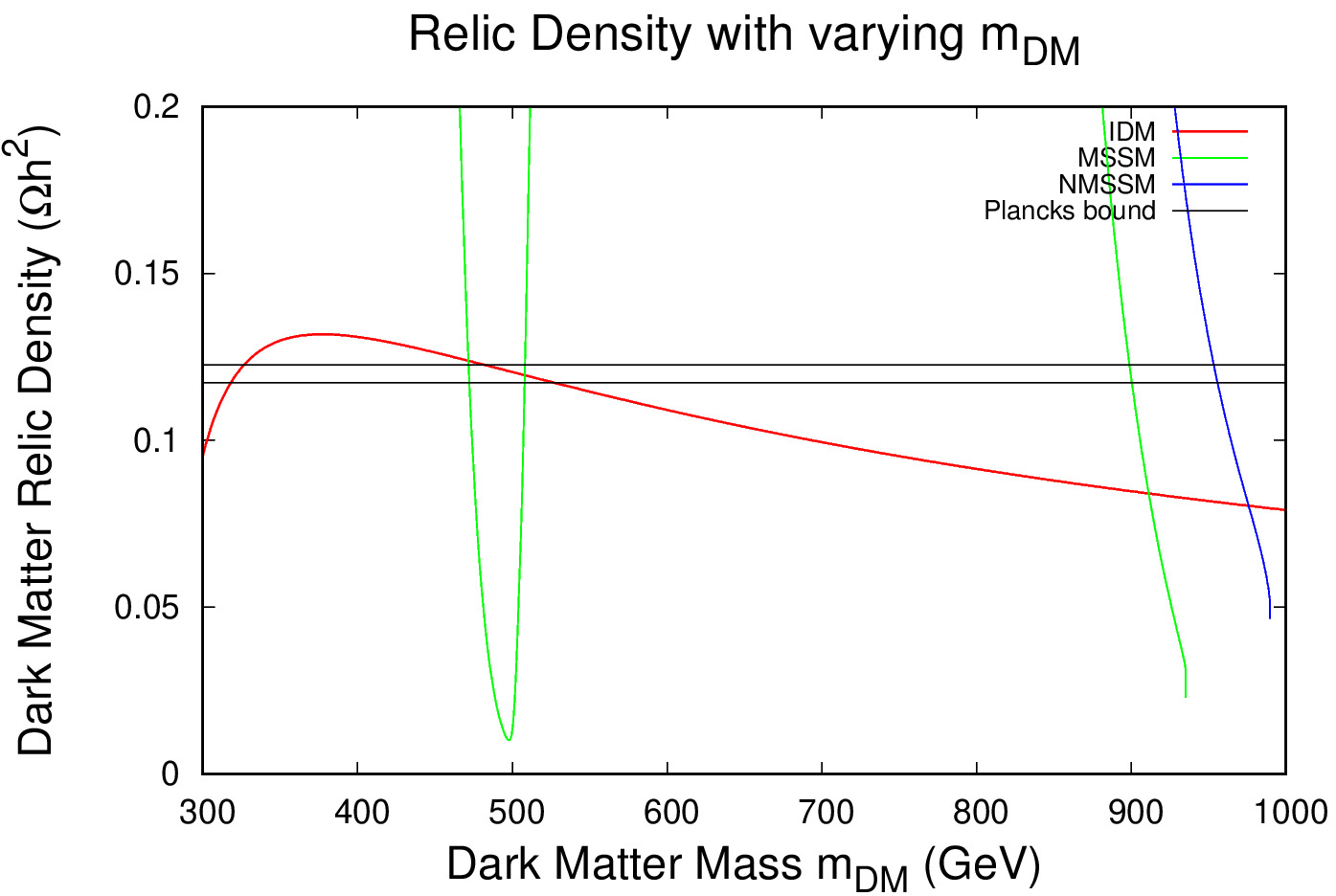}
	\caption{Zoomed-in view of relic densities at heavy masses, highlighting the viable mass windows (shaded green band) for each model consistent with Planck bounds.}
	\label{fig:viable_windows}
\end{figure}

This analysis demonstrates that the relic density profile with its trends, resonance structures, and viable mass windows, serves as a useful discriminatory tool. The IDM is characterized by a strong Higgs resonance and broad viable regions. The MSSM is marked by a pronounced slepton-mediated dip, while the NMSSM shows diluted features due to its singlet extension. 

These distinct signatures provide a practical basis for differentiating between doublet and supersymmetric dark matter origins in future phenomenological studies and experimental searches. These benchmark points are the only viable regions in the whole parameter space of Table I, and will be compared for fine-tuning requirements and tested against Planck's data to check for Cosmic Microwave Background energy injections for compatibility with bounds from observational cosmology \cite{Planck2015}.

The heavy WIMP parameter space identified for the IDM remains theoretically consistent: with modest quartic couplings ($\lambda_L \sim 0.001$) and TeV-scale scalar masses, the model maintains perturbativity and vacuum stability up to high scales. Previous studies of the IDM~\cite{IDMHiggsLHC} indicate that such parameter points can remain valid up to the Planck scale without encountering Landau poles or vacuum instability. This is further reinforced by \textit{MicrOEMGAs}, which does not indicate any such inconsistencies during the scans. This theoretical robustness, combined with the broad phenomenological viability, further distinguishes the IDM from SUSY frameworks where viable regions often correspond to specific, fine-tuned points in a more constrained parameter space.
 
Despite predicting similar relic abundance, the freeze-out dynamics are different for each model and further add to the discriminatory features of these frameworks.

\subsection*{Fine-tuning Comparison}

The fine-tuning measure $\Delta_{M_{\rm DM}}$ quantifies the sensitivity of the dark matter relic density to variations in the dark matter mass. Smaller values of $\Delta$ indicate that the relic density is less sensitive to parameter changes, implying a more “natural” model. 

The fine-tuning metric calculations have been carried out on the viable mass regions from FIG. 7. For the Inert Doublet Model (IDM), we compute $\Delta_{M_{\rm DM}}$ across all Planck-allowed points in two viable mass regions. The resulting curves are shown in Fig. 7 (region 1) and Fig. 8 (region 2). In region 1, $\Delta$ ranges approximately from 1.4 to 1.9, while in region 2 it ranges from 0.47 to 0.52, indicating very low fine-tuning across the allowed parameter space.

\begin{figure}[H]
	\centering
	\includegraphics[width=0.7\linewidth]{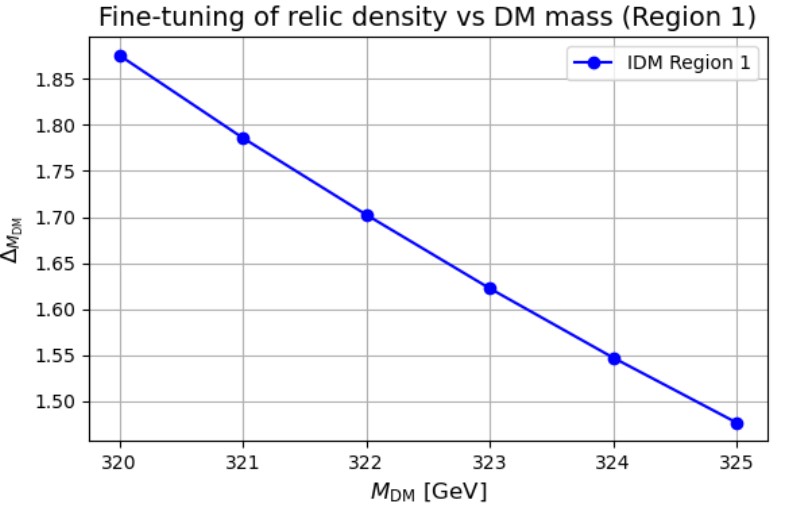}
	\caption{The fine-tuning metric $\Delta_{M_{\rm DM}}$ shows an approximately linear decreasing trend for viable region 1 of the IDM ($320-325$ GeV). }
	\label{fig:fig7}
\end{figure}

\begin{figure}[H]
	\centering
	\includegraphics[width=0.7\linewidth]{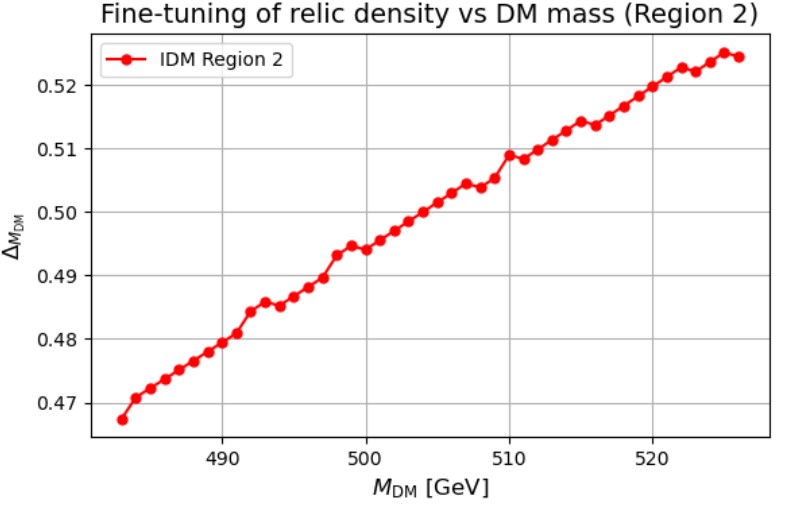}
	\caption{The fine-tuning metric $\Delta_{M_{\rm DM}}$ shows an increasing trend for viable region 2 of the IDM ($482-526$ GeV).}
	\label{fig:fig8}
\end{figure}

The absolute value was used in Eq. 13, since the sign is an irrelevant artifact that depicts whether the relic density itself is increasing or decreasing in the viable region. Similar computations were carried out for the SUSY models which yielded far narrower viable regions (FIG. 7).

So for the MSSM and NMSSM, only a few points satisfy the Planck relic density bounds. Table II lists the estimated $\Delta_{M_{\rm DM}}$ values at points closest to the Planck central value. The MSSM shows high fine-tuning, with $\Delta$ ranging from 26 to 87.6, while the NMSSM exhibits moderate fine-tuning, with $\Delta \sim 17.6$ albeit with only one viable mass region.

\begin{table}[H]
	\centering
	\caption{Estimated fine-tuning $\Delta_{M_{\rm DM}}$ for MSSM and NMSSM. Only Planck-allowed or closest-to-Planck points are considered.}
	\begin{tabular}{|l|c|c|}
		\hline
		Model & $M_{\rm DM}$ [GeV] & $\Delta_{M_{\rm DM}}$ \\
		\hline\hline
		MSSM & 471.63 & 42.127 \\
		\hline
		MSSM & 508.46 & 87.6 \\
		\hline
		MSSM & 899.714 & 26.257 \\
		\hline
		NMSSM & 954.07 & 17.593 \\
		\hline
		NMSSM & 954.74 & 17.580 \\
		\hline
	\end{tabular}
	\label{tab:SUSY_Delta}
\end{table}

Averaging over Planck-allowed points, the IDM exhibits $\langle\Delta\rangle_{IDM} \sim 1.63$ across its combined 44 GeV wide viable regions, while MSSM and NMSSM show much higher sensitivity, with average $\langle\Delta\rangle_{MSSM} \sim 52$ and a combined viability width of 2.4 GeV. NMSSM exhibited a viability region of 3.4 GeV with $\langle\Delta\rangle_{NMSSM} \sim 17.6$.

\subsection*{Freeze-out Quantities for Viable Regions}

For all models considered, we evaluate the freeze-out parameter $X_f$, freeze-out temperature $T_f$, and the corresponding freeze-out redshift $z_f$ for parameter points satisfying the Planck relic density constraint in Eq. 12.

In the IDM, we perform a continuous scan over the dark matter mass in the heavy regime,
$300~\mathrm{GeV} \le m_{\rm DM} \le 1000~\mathrm{GeV}$,
with a fixed inert scalar mass splitting to allow co-annihilation.
Within the Planck-allowed regions, the freeze-out parameter remains remarkably stable,
\begin{equation}
	X_f \simeq 26.1\text{--}26.6,
\end{equation}
exhibiting only a mild dependence on the dark matter mass.

The freeze-out temperature increases monotonically with $m_{\rm DM}$, ranging from
$T_f \simeq 12~\mathrm{GeV}$ at $m_{\rm DM}\sim 320~\mathrm{GeV}$ to
$T_f \simeq 20~\mathrm{GeV}$ near $m_{\rm DM}\sim 520~\mathrm{GeV}$.
The corresponding freeze-out redshift lies in the range
\begin{equation}
	z_f \sim (1.6\text{--}2.6)\times10^{14},
\end{equation}
reflecting the thermal decoupling of dark matter deep in the radiation-dominated era. FIG. 10 and FIG. 11 shows these quantities for the 320-325 GeV, and 482-526 GeV viable regions, respectively.

\begin{figure}[H]
	\centering
	\includegraphics[width=0.68\linewidth]{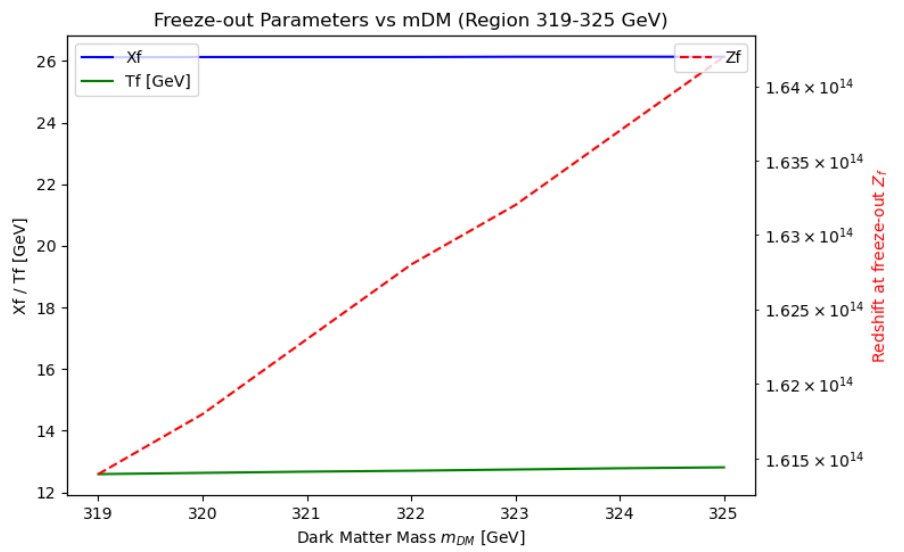}
	\caption{$X_f$ (blue), $T_f$ (green) and $z_f$ (dotted red) plotted against $m_{DM}$ from 320 GeV to 325 GeV.}
	\label{fig:fig9}
\end{figure}

\begin{figure}[H]
	\centering
	\includegraphics[width=0.68\linewidth]{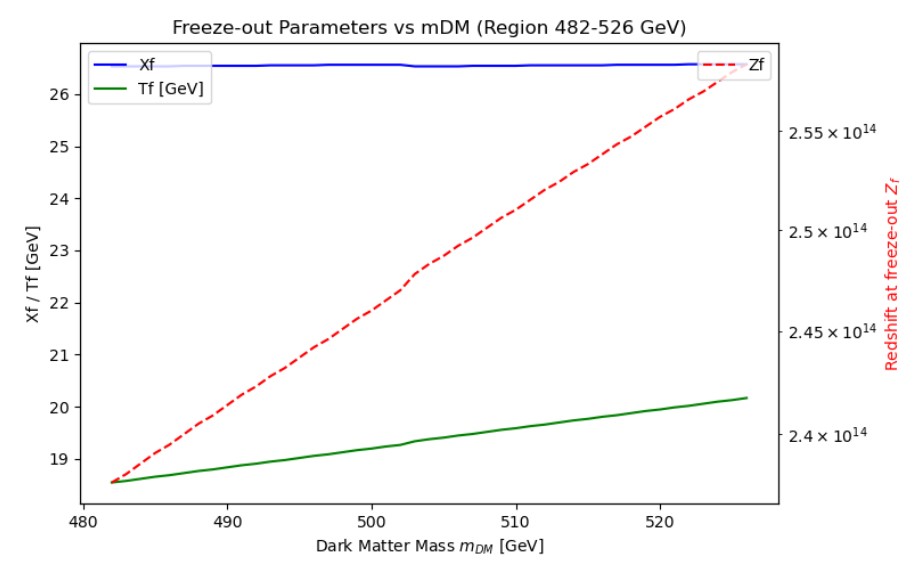}
	\caption{$X_f$ (blue), $T_f$ (green) and $z_f$ (dotted red) plotted against $m_{DM}$ from 482 GeV to 526 GeV.}
	\label{fig:fig10}
\end{figure}

The smooth behavior of $X_f$, $T_f$, and $z_f$ across the viable IDM region indicates that freeze-out occurs under nearly identical cosmological conditions throughout the heavy-mass regime.

In contrast, viable relic-density regions in the MSSM and NMSSM are confined to narrow windows of the gaugino mass parameter.
As a result, the corresponding freeze-out quantities exhibit negligible variation within each allowed region.
For representative MSSM points, we find
\begin{equation}
	X_f \simeq 18.1\text{--}32.5, \qquad
	z_f \sim (2.3\text{--}4.2)\times10^{14},
\end{equation}
while the NMSSM benchmark yields
\begin{equation}
	X_f \simeq 28.6, \qquad
	z_f \simeq 4.3\times10^{14}.
\end{equation}

Due to the narrowness of these viable regions, the freeze-out quantities for the MSSM and NMSSM are summarized in tabular form rather than as continuous scans in Table III below.

\begin{table}[H]
	\centering
	\caption{Viable MSSM and NMSSM points: Relic density and freeze-out parameters}
	\begin{tabular}{|c|c|c|c|c|}
		\hline
		Model  & $m_{DM}$ [GeV]  & $X_f$ & $T_f$ [GeV] & $Z_f$ \\
		\hline\hline
		MSSM  & 508.465  & 24.55 & 20.71 & $2.654\times10^{14}$ \\
		\hline
		MSSM  & 471.929  & 26.03 & 18.13 & $2.323\times10^{14}$ \\
		\hline
		MSSM  & 472.028  & 26.04 & 18.13 & $2.323\times10^{14}$ \\
		\hline
		MSSM  & 899.5    & 27.69 & 32.48   & $4.16\times10^{14}$ \\
		\hline
		NMSSM & 954.75   & 28.58 & 33.40 & $4.280\times10^{14}$ \\
		\hline
	\end{tabular}
	\label{tab:susy_relic}
\end{table}

This highlights a qualitative distinction between the IDM and supersymmetric scenarios as all models register different freeze-out parameters, each with their own imprint on the thermal history of the Universe, despite yielding the same relic abundance for dark matter.

Although the freeze-out redshift is not a directly measurable cosmological observable, it provides a useful bookkeeping parameter that links particle-physics model parameters to the thermal history of the early Universe.

Across all models, freeze-out occurs well before the electroweak phase transition and deep in the radiation-dominated era, validating the standard assumptions underlying the relic density calculation.

\subsection*{CMB Constraints on Viable Dark Matter Regions}

In this section, we present the results for the effective annihilation parameter $p_{\mathrm{ann}}$ across the relic-density–allowed regions of the Inert Doublet Model (IDM) and supersymmetric scenarios. All parameter points shown satisfy the Planck 2018 relic density constraint, and are further tested against CMB energy-injection limits.

FIG. 12 displays the behaviour of $p_{\mathrm{ann}}$ for the IDM region spanning $m_{\mathrm{DM}} \in [318, 326]~\mathrm{GeV}$. Across this interval, the annihilation parameter remains nearly constant at $\mathcal{O}(10^{-28})~\mathrm{cm^3\,s^{-1}\,GeV^{-1}}$, with a mild monotonic decrease as the dark matter mass increases.

\begin{figure}[H]
	\centering
	\includegraphics[width=0.7\linewidth]{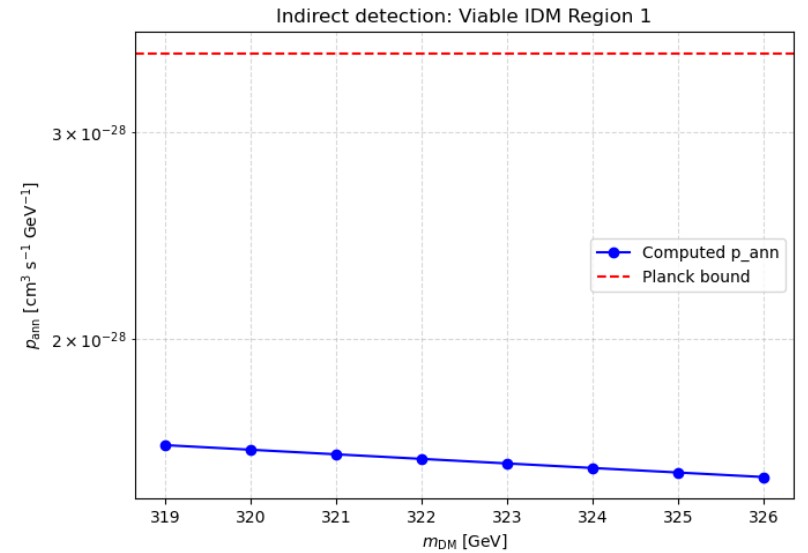}
	\caption{The annihilation parameter $p_{\mathrm{ann}}$ for the first viable IDM region in linear scale.}
	\label{fig:fig11}
\end{figure}

We note that our analysis assumes a constant efficiency factor \( f_{\text{eff}} = 0.3 \), which is representative of weak-scale dark matter annihilating dominantly into electroweak final states. In reality, \( f_{\text{eff}} \) depends on the specific annihilation channels and dark matter mass, and can vary by a factor of \( \sim 2-3 \) across different final states. However, given that all our viable points yield \( p_{\rm ann} \) values at least an order of magnitude below the Planck limit, all points lie comfortably below the Planck upper bound on $p_{\mathrm{ann}}$, indicating that this region is fully consistent with CMB constraints. Combined with its agreement with the observed relic density, this mass window represents a phenomenologically viable and cosmologically safe dark matter regime within the IDM framework.

\begin{figure}[H]
	\centering
	\includegraphics[width=0.7\linewidth]{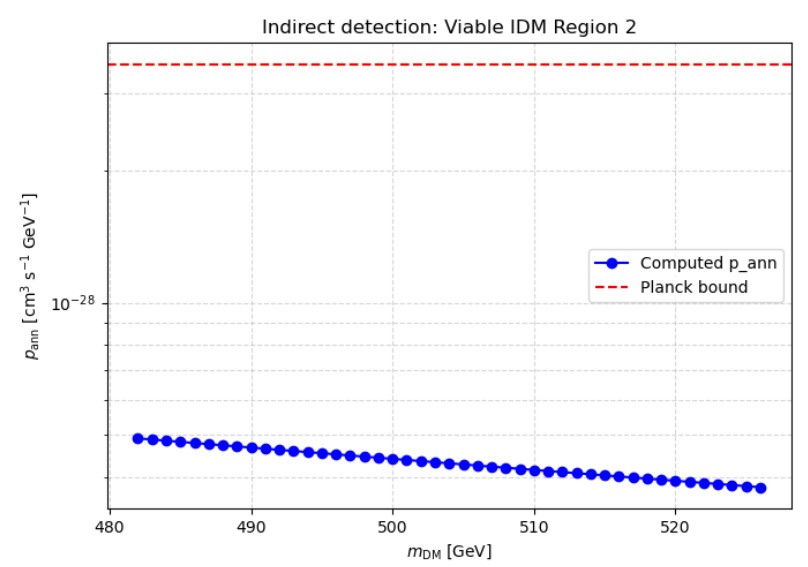}
	\caption{The annihilation parameter $p_{\mathrm{ann}}$ for the first viable IDM region in logarithmic scale.}
	\label{fig:fig12}
\end{figure}

The second IDM mass window, shown in FIG. 13 corresponds to $m_{\mathrm{DM}} \in [482, 526]~\mathrm{GeV}$. In this region, $p_{\mathrm{ann}}$ is systematically suppressed relative to the low-mass case, taking values of order $10^{-29}~\mathrm{cm^3\,s^{-1}\,GeV^{-1}}$.

This suppression arises primarily from the inverse dependence of $p_{\mathrm{ann}}$ on the dark matter mass, while the annihilation cross section remains close to its canonical thermal value. As a result, the entire region lies more than an order of magnitude below the Planck limit, rendering it robustly unconstrained by current CMB observations.

FIG. 14 presents the corresponding results for the MSSM and NMSSM benchmark points satisfying the relic density constraint. For the MSSM, viable solutions around $m_{\mathrm{DM}} \simeq 470~\mathrm{GeV}$ and $m_{\mathrm{DM}} \simeq 510~\mathrm{GeV}$ yield $p_{\mathrm{ann}}$ values ranging from $\mathcal{O}(10^{-30})$ to $\mathcal{O}(10^{-29})~\mathrm{cm^3\,s^{-1}\,GeV^{-1}}$. Heavier MSSM solutions near $m_{\mathrm{DM}} \simeq 900~\mathrm{GeV}$ further suppress $p_{\mathrm{ann}}$ to $\mathcal{O}(10^{-31})~\mathrm{cm^3\,s^{-1}\,GeV^{-1}}$.

Similarly, the NMSSM benchmark at $m_{\mathrm{DM}} \simeq 955~\mathrm{GeV}$ exhibits a comparably small annihilation parameter, well below the Planck bound. In all supersymmetric cases considered, CMB constraints do not impose additional restrictions beyond those already enforced by relic density considerations.

\begin{figure}[H]
	\centering
	\includegraphics[width=0.7\linewidth]{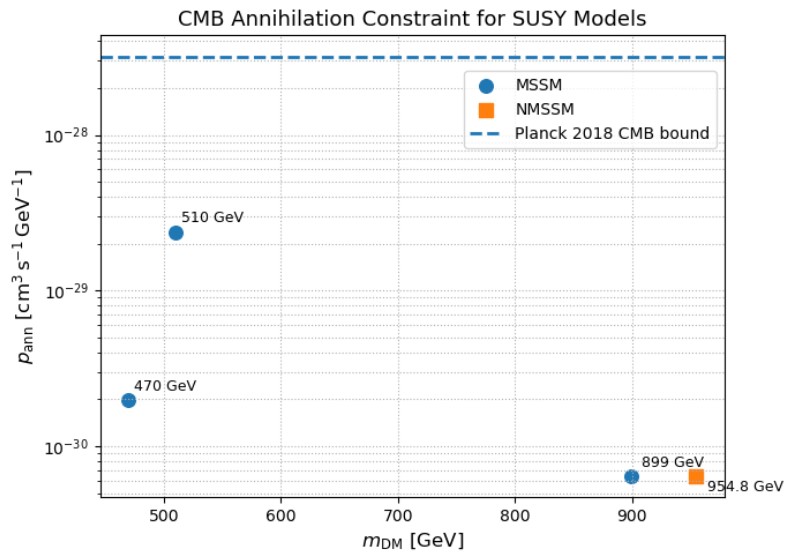}
	\caption{The annihilation parameter $p_{\mathrm{ann}}$ yielded by the viable points of MSSM and NMSSM compared the Planck's CMB bound.}
	\label{fig:fig13}
\end{figure}

Across all models and mass ranges studied, the effective annihilation parameter $p_{\mathrm{ann}}$ remains safely below the Planck 2015 upper limit. This demonstrates that the relic-density–compatible regions of the IDM, MSSM, and NMSSM considered in this work are also consistent with CMB energy-injection constraints. Consequently, these scenarios represent cosmologically viable dark matter candidates within current observational sensitivity. 

The fact that all relic-density–viable regions also comfortably satisfy CMB constraints demonstrates that the identified parameter spaces are not only phenomenologically interesting but also cosmologically robust, offering complete candidate profiles for future testing.\\

\section{Conclusion}

This study has established a multi-faceted, relic-density-based framework for discriminating between scalar doublet and supersymmetric dark matter candidates. Through controlled, unified parameter scans of the Inert Doublet Model (IDM), the Minimal Supersymmetric Standard Model (MSSM), and the Next-to-Minimal Supersymmetric Standard Model (NMSSM) using \textsc{MicrOMEGAs}, we have identified distinct phenomenological fingerprints for each paradigm.

Our analysis reveals that the relic density \textit{topology} serves as a primary discriminant. The IDM is characterized by a broad ``viability plateau'' in the heavy-mass regime, reflecting efficient gauge-driven annihilation and remarkably low parametric fine-tuning ($\Delta_{M_{\text{DM}}} \sim 0.5\text{--}1.9$). In stark contrast, the SUSY frameworks attain the correct relic density only through narrow resonances or mass windows, exhibiting fine-tuning that is greater by one to two orders of magnitude ($\Delta_{M_{\text{DM}}} \sim 18\text{--}88$). A key signature emerges in the heavy mass regime: a sharp slepton-mediated annihilation dip near $500~\mathrm{GeV}$ in the MSSM that is uniquely damped in the NMSSM due to singlino admixture, providing a clear channel-specific fingerprint to tell the two SUSY models apart.

These topological and fine-tuning features are complemented by their cosmological signatures. The viable regions for all models are consistent with Cosmic Microwave Background constraints, and their freeze-out histories—stable and predictable for the IDM versus pinpointed to specific epochs for the SUSY cases—add another layer of distinction.

Collectively, these results provide a concrete diagnostic toolkit. The discovery of a heavy thermal WIMP would favor a scalar doublet origin like the IDM if its relic density is measured across a broad mass range without sharp resonances. Conversely, evidence pointing to a narrow viable mass window or a specific annihilation channel signature would point toward a supersymmetric origin, with the detailed resonance structure further discriminating between MSSM and NMSSM scenarios.

Future work can directly integrate this framework with experimental data. Applying it to global fits that include direct and indirect detection limits, as well as projecting these discriminants onto collider observables, will sharpen its utility in guiding the interpretation of the next generation of dark matter searches.

\end{document}